# Hydrogen-rich hydrate at high pressures up to 104 GPa

Alexander F. Goncharov[1], Elena Bykova[2], Iskander Batyrev [3], Maxim Bykov[4], Eric Edmund[1], Amol Karandikar[1], Mahmood Mohammad[5], Stella Chariton[6], Vitali Prakapenka[6], Konstantin Glazyrin[7], Mohamed Mezouar[8], Gaston Garbarino[8], Jonathan Wright[8]

[1] Earth and Planets Laboratory, Carnegie Science, Washington, DC 20015, USA
[2] Goethe-Universität Frankfurt am Main, Facheinheit Mineralogie, 60438 Frankfurt am Main, Germany
[3] U.S. Army DEVCOM Army Research Laboratory, FCDD-RLA-WA, Aberdeen Proving Ground, Maryland, 21005 U.S.A.
[4] Institute of Inorganic and Analytical Chemistry, Goethe University Frankfurt, Max-von-Laue-Straße 7, 60438 Frankfurt am Main, Germany
[5] Department of Mathematics, Howard University, Washington DC 20059 USA.
[6] Center for Advanced Radiation Sources, The University of Chicago, Chicago, Illinois 60637, USA
[7] Deutsches Elektronen-Synchrotron DESY, Notkestr. 85, 22607 Hamburg
[8] European Synchrotron Radiation Facility BP 220, 38043 Grenoble Cedex, France

**Gas hydrates are considered fundamental building blocks of giant icy planets like Neptune and similar exoplanets. The existence of these materials in the interiors of giant icy planets, which are subject to high pressures and temperatures, depends on their stability relative to their constituent components. In this study, we reexamine the structural stability and hydrogen content of hydrogen hydrates, $(H_2O)(H_2)_n$, up to 104 GPa, focusing on hydrogen-rich materials. Using synchrotron single-crystal X-ray diffraction, Raman spectroscopy, and first-principles theoretical calculations, we find that the $C_2$-filled ice phase undergoes a transformation to a $C_3$-filled ice phase over a broad pressure range of 47–104 GPa at room temperature. The $C_3$ phase contains twice as much molecular $H_2$ as the $C_2$ phase. Heating the $C_2$-filled ice above approximately 1500 K induces the transition to the $C_3$ phase at pressures as low as 47 GPa; upon decompression, this phase remains metastable down to 40 GPa. These findings establish new stability limits for hydrates, with implications for hydrogen storage and the interiors of planetary bodies.**

## I. INTRODUCTION

Under high pressure, water and hydrogen are known to form various stoichiometric compounds with distinct compositions. At low pressures, the well-known clathrate sII crystallizes. At higher pressures, up to 3 GPa, several compounds— $C_{-1}$, $C_0$, $C_1$, $C_2$ –have been reported [1-8]. These compounds are characterized as filled ices, where $H_2O$ host molecules form a sublattice resembling one of the pure ice phases, while $H_2$ molecules occupy interstitial positions or may even replace $H_2O$ molecules within the prototype ice structure.

The high-pressure $C_2$-filled ice can be viewed as a cubic structure similar to ice VII—a high-pressure form of ice stable above 2 GPa— in which one of the $H_2O$ sublattices is replaced by $H_2$ molecules [6]. Consequently, the $C_2$ phase has a composition of $(H_2O)H_2$. This phase has been reported to persist up to 80 GPa [7, 9], undergoing structural changes between 35–40 GPa and 55–60 GPa, hypothesized to correspond to hydrogen bond symmetrization and the formation of a denser structure, respectively. However, theoretical calculations suggest that the $C_2$ phase is thermodynamically unstable above 19 GPa [10]. These calculations also predict that the $C_2$ structure should transition to a tetragonal or orthorhombic form above 20 GPa [10].

Another filled ice structure, $C_3$, with an increased hydrogen composition of $(H_2O)(H_2)_2$, has been theoretically predicted at high pressures [10]. This phase, expected to stabilize above 30 GPa in hydrogen-rich conditions, begins as a cubic structure with an ice sublattice identical to that of $C_2$. However, the $H_2$ molecules in $C_3$ are predicted to occupy different crystallographic positions compared to those in $C_2$, resulting in twice the number of $H_2$ molecules. The transformation between $C_2$ and $C_3$ was recently reported in the 44–60 GPa range, facilitated by laser heating to 1100 K, based on experiments conducted under hydrogen-poor conditions [11]. However, structural information on $C_3$ remains limited, and both the hydrogen atom positions, and the overall composition have so far been inferred primary by comparing the experimental observations with first-principles calculations.

Here, we report experiments conducted under hydrogen-rich conditions, demonstrating that the $C_2$-to-$C_3$ transition can occur even at room temperature, although it progresses gradually over a broad pressure range of 47–103 GPa. Within this range, the hydrogen content steadily increases until it reaches the stoichiometric ratio. Laser heating experiments at 50–69 GPa reveal an abrupt transformation in the quenched sample, evidenced by an increase in volume and a softening of the main intramolecular stretching mode (vibron) of the $H_2$ molecule. At 69 GPa, synchrotron single-crystal X-ray diffraction (SC XRD) determines that the structure of the $C_3$ phase is cubic, with the space group $Fd\overline{3}m$. The positions of the $H_2$ molecular centers and the hydrogen atoms in the $H_2O$ framework were obtained.

First-principles theoretical calculations indicate that the $C_2$ phase remains stable up to 14 GPa, while the $C_3$ phase becomes stable above 23 GPa, as approximated by the hydrogen-ordered $Pna2_1$ and $P4_1$ structures. Both phases are predicted to be metastable in the intermediate pressure range of 14-23 GPa. The calculated pressure-dependent volume changes of these phases are consistent with experimental observations. Furthermore, Raman spectra calculated over the 5-100 GPa range show a softening of the main hydrogen vibron mode associated with the $C_2$ to $C_3$ transition, in qualitative agreement with the experimental results.

## II. MATERIALS AND METHODS

### A. Experiments

X-ray diffraction and Raman spectroscopy experiments (see Table S1 of Supplemental Materials [12]) were conducted using diamond anvil cells (DACs) optimized for single-crystal X-ray diffraction with culet diameters of 200 μm and 100 μm. Pressure measurements were performed using small chips of gold and ruby, which served as pressure markers via X-ray diffraction and optical fluorescence, respectively. Rhenium gaskets were used, and a hole with a diameter of 60–100 μm was prepared for the initial loading of ultrapure water and pressure sensors. To create the conditions necessary for synthesizing hydrogen-rich compounds at high pressures, an air vesicle was introduced into the water, with a radial dimension of 0.4–0.7 times the diameter of the gasket hole. The DAC was subsequently gas-loaded with compressed $H_2$ gas at 0.14 GPa in the Earth and Planets Laboratory (EPL) of Carnegie Institution for Science, producing $H_2$-$H_2O$ mixtures with sufficiently high hydrogen content to facilitate the formation of high-pressure compounds.

Small flakes and disks of metallic Au were positioned in the DAC cavity to absorb heat from an infrared laser (1064 nm) during laser heating experiments conducted at GSECARS (Sector 13 of the Advanced Photon Source) and EPL. Additionally, a $CO_2$ laser (10.6 μm) was used at EPL to directly heat the samples by via absorption by their vibrational modes. Due to the unstable nature of heating, spectradiometric temperature measurements were not feasible. However, the sample temperature was estimated to be up to 1500 K based on brief visual observation of transient thermal radiation.

Synchrotron XRD investigations were conducted at GSECARS, the Extreme Conditions Beamline (ECB) at Petra III (DESY, Hamburg), and at ESRF beamlines ID27 [13] and ID11. These experiments utilized a tightly focused (<3 μm in diameter) monochromatic beam with wavelengths of 0.2952 Å and 0.3344 Å (GSECARS), 0.2919 Å (ECB), 0.3738 Å (ID27) and 0.2846 Å (ID11), recorded using position-sensitive 2D array detectors. Powder diffraction maps were measured across the entire high-pressure cavity with a radial step of 1-4 μm. Regions of interest, identified by the detection of single-crystal Bragg reflections over a broad range of the moment transfer Q, were subsequently analyzed using SC XRD technique to examine the sample structure in the corresponding areas. All XRD measurements were performed at room temperature. Raman measurements (including mapping) were performed with custom made Raman systems at GSECARS [14], EPL [15], and WiTec spectrometers at EPL and Goethe University Frankfurt using excitation wavelengths of the 488, 532, and 660 nm.

The procedure for multigrain SC XRD crystallography and structure determination have been detailed in previous publications (e.g., Refs. [16, 17]). During single-crystal XRD measurements, diffraction images were recorded as the samples were rotated around a vertical ω-axis in a range of ±35° with an angular step Δω = 0.5° and an exposure time of 1-2 seconds per frame. The CrysAlisPro software package [18] along with the DAFi module, which identifies subsets of reflections from individual domains within a complete set of SC-XRD data [19], was used to analyze the single-crystal diffraction data, enabling SC XRD analysis for multigrain samples. To calibrate an instrumental model in the CrysAlisPro software —encompassing parameters such as the sample-to-detector distance, detector origin, goniometer angle offsets, and the rotation of the X-ray beam and the detector around the instrument axis —a single crystal of orthoenstatite (($Mg_{1.93}Fe_{0.06}$)($Si_{1.93}$, $Al_{0.06}$)$O_6$, *Pbca* space group, $a = 8.8117(2)$, $b = 5.1832(1)$, and $c = 18.2391(3)$ Å) was used.

## B. Theoretical calculations
First-principles calculations were conducted within the framework of density functional theory (DFT) [20, 21], using the Perdew–Burke–Ernzerhof (PBE) [22] exchange-correlation functional and the plane wave/pseudopotential approach as implemented in the CASTEP simulation package [23]. For band gap calculations, results were compared with those obtained using hybrid three-parameter BLYP functional [24], which balances computational efficiency with accuracy. The parameters defining the weights of Hartree–Fock exchange and DFT exchange-correlation contributions were empirically determined through a linear least-squares fit to experimentally determined energies.

Norm-conserving pseudopotentials from the CASTEP database were employed, combined with plane waves up to a kinetic energy cutoff of 770 eV. CASTEP calculates finite basis set corrections by evaluating the total energy at several cutoff energies and numerically computing the derivative required for the correction term. Typically, three reference points are sufficient for such calculations. For Brillouin zone integrations, a Monkhorst–Pack grid was used, ensuring a grid point spacing of less than 0.03 Å$^{-1}$.

The convergence criteria for geometry optimization were set to ensure high accuracy, including an energy change of less than $5 \times 10^{-8}$ eV per atom between steps, a maximum force of less than 0.005 eV Å$^{-1}$, and a maximum stress tensor component of less than 0.001 GPa. Grimme dispersion corrections were applied to account for van der Waals interactions [25].

Phonon frequencies were calculated using density functional perturbation theory (DFPT) [26]. Phonon dispersion and Raman spectra were determined using the linear response method, as implemented in the CASTEP code [23, 26].

## III. RESULTS AND DISCUSSIONS

Consistent with previous studies, our experiments demonstrate that the filled ice $C_2$ phase forms above 6 GPa, as indicated by low-angle diffraction peaks that are absent in ice VII. Notably, ice VII is consistently observed in the high-pressure cavity, even under hydrogen-rich conditions. The XRD patterns of the $C_2$ phase correspond to a cubic $Fd\bar{3}m$ structure [6] at low pressures. However, above 24 GPa, the Bragg peaks broaden and, in some cases, split (see also Ref. [7]), indicating a reduction in symmetry (Fig. 1 and Fig. S1 of the Supplemental Materials [12]).

The observed XRD patterns can be satisfactorily described by the theoretically predicted $Pna2_1$ structure [10] up to 64 GPa (Fig. S1 of the Supplemental Materials [12]) as evidenced by the good agreement between the experimentally observed Bragg reflections and the calculated peak positions. At pressures exceeding 64 GPa, the patterns become increasingly complex due to the emergence of the $C_3$ phase (Fig. 1). The volume of the $Pna2_1$ structure, fitted to our experimental data, closely aligns with the value determined from the first diffraction peak under the assumption of cubic symmetry. However, the experimental peak splitting—and consequently, the deviation of the structure from cubic symmetry—is significantly smaller than predicted theoretically (Fig. S2 of the Supplemental Materials [12]).

This discrepancy likely arises from differences in the treatment of hydrogen ordering effects in theoretical versus experimental approaches. Theoretical models generally focus on hydrogen-ordered structures for computational tractability, representing idealized configurations. In contrast, experiments conducted at 300 K inherently reflect hydrogen-disordered states, as thermal energy at this temperature disrupts hydrogen ordering and introduces dynamic disorder. These differences in theoretical and experimental conditions can result in discrepancies in predicted and observed structural properties.

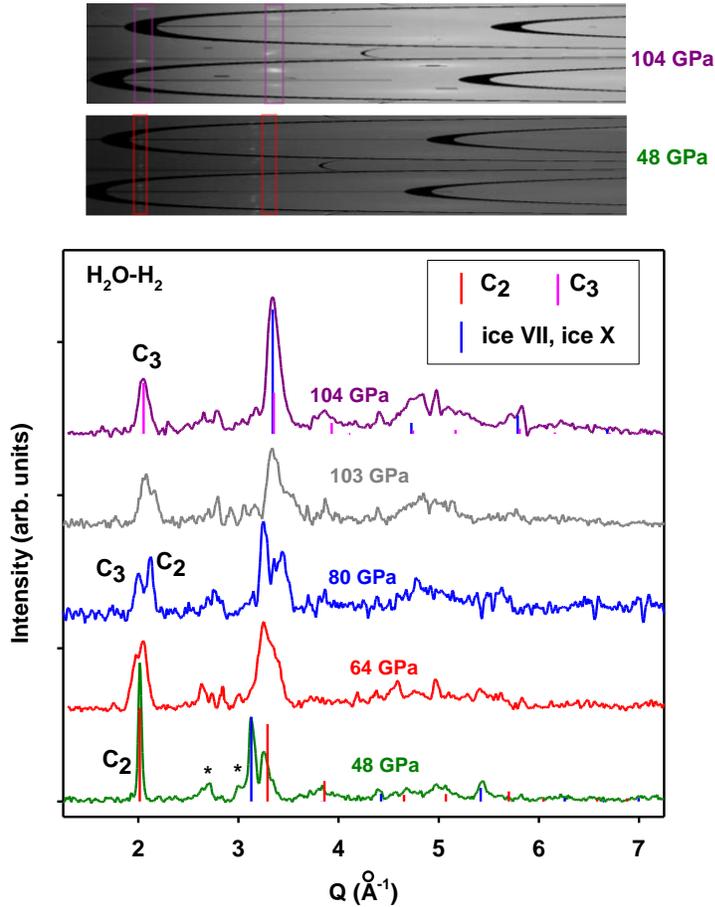

**Figure 1.** X-ray diffraction patterns measured during pressure increases at 300 K at GSECARS. The scattering vector, $\boldsymbol{Q}$, is defined as $Q = 4\pi \sin(\theta)/\lambda$, where $\lambda$ is the X-ray wavelength and $2\theta$ is the diffraction angle. The top two panels display 2D diffractograms in rectangular coordinates for 48 and 104 GPa; the characteristic reflections of $C_2$ and $C_3$ phases are boxed using the color scheme of the main panel. Vertical lines mark the calculated diffraction line positions for the $Pna2_1$ $C_2$ phase at 48 GPa and the $Fd\bar{3}m$ $C_3$ phase at 104 GPa. Asterisks indicate Bragg peaks from the rhenium gasket. The X-ray wavelength was 0.2952 Å for 64 and 104 GPa, and 0.3344 Å for 48, 80, and 103 GPa.

X-ray diffraction patterns of filled ice measured during pressure increases above 50 GPa reveal a gradual transformation (Fig. 1), marked by the splitting of the first diffraction peak. A new, lower-angle peak emerges and becomes more pronounced with increasing pressure. At 103 GPa, after maintaining the sample at this pressure for several weeks in the presence of excess $H_2$, the diffraction peak corresponding to the $C_2$ phase dropped below the detection limit. The filled ice observed in the chamber is well described by an $Fd\bar{3}m$ cubic phase, which exhibits a larger volume compared to the initial $C_2$ phase. This volume closely matches values extrapolated from lower-pressure data for the C₃ phase (Ref. [11]) and from our laser-heating experiments, as detailed below.

Two types of laser-heating experiments were performed to accelerate the transformation at pressures ranging from 30 to 86 GPa (Table S1 of the Supplemental Materials [12]). In some experiments, thin gold flakes and focused ion beam (FIB)-machined gold discs (<3 μm thick and approximately 50 μm in diameter) with seven symmetrically drilled small holes were placed in the high-pressure cavity (Fig. S3 of the Supplemental Materials [12]). These served as absorbers for near-infrared laser light (1.07 μm), enabling localized heating of the sample near the absorber. Additionally, an infrared $CO_2$ laser (10.6 μm) was used to heat the bulk of the inclusion compound by coupling with low-frequency translational and rotational lattice excitations.

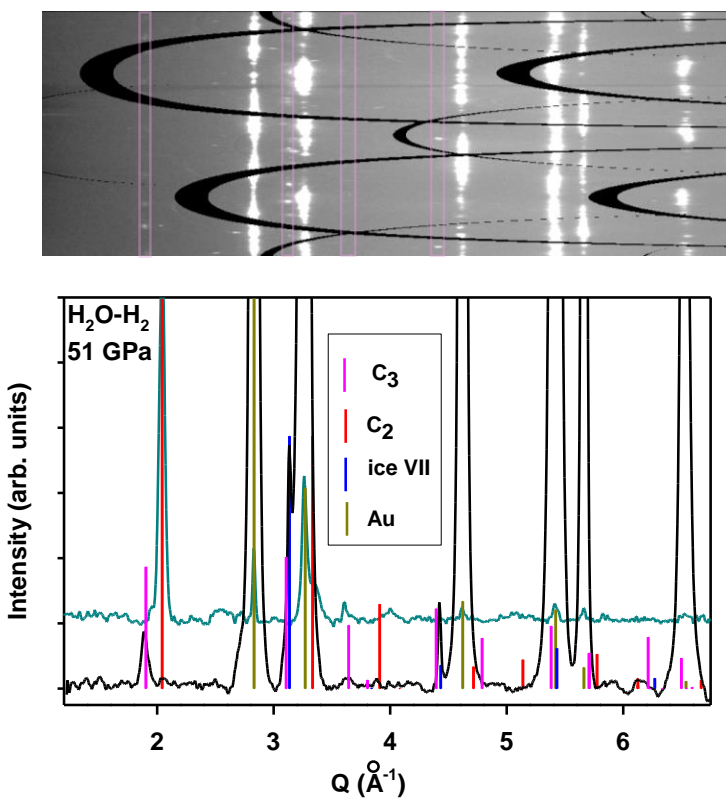

**Figure 2.** X-ray diffraction patterns measured after near-IR laser heating of the $C_2$ compound at 51 GPa. The top panel shows a 2D diffractogram at the $C_3$ position in rectangular coordinates, with the characteristic reflections of the $C_3$ phase boxed. Vertical lines in the main panel indicate the calculated diffraction line positions for the $Fd\overline{3}m$ $C_2$ and $Fd\overline{3}m$ $C_3$ phases, $Fm\overline{3}m$ Au, and $Pn\overline{3}m$ $H_2O$ ice VII. The XRD signal from the Au coupler is particularly strong in the bottom pattern, which was measured in close proximity to the coupler. The X-ray wavelength used was 0.3738 Å.

Near-IR heating at 30 GPa did not produce noticeable changes in the XRD patterns. However, significant changes were observed during near-IR heating at 47–57 GPa using FIB-machined Au couplers. In two experiments conducted under these conditions, where the maximum temperature remained below 1500 K, very thin sleeves of the $C_3$ phase formed and were clearly detected through XRD mapping (Fig. 2 and Fig. S3 of the Supplemental Materials [12]). The XRD diagnostics of the $C_2 - C_3$ transition were unambiguous: in the transformed sample, the Bragg peaks shifted significantly toward lower angles (Fig. 2), indicating an increase in lattice parameters, and, thus, in the unit cell volume, consistent with the transition from the $C_2$ to the $C_3$ phase.

IR $CO_2$ laser heating at 52 GPa caused the pressure within the cavity to increase to 69 GPa and led to the formation of a significant amount of the $C_3$ phase, as evidenced by an increase in the unit cell volume, as described below. Visual observations revealed the formation of a vesicle in the center of the high-pressure cavity following laser heating (Fig. S4 of the Supplemental Materials [12]). Powder XRD patterns (Fig. S5 of the Supplemental Materials [12]) confirm the formation of the $C_3$ phase after heating, localized within the central vesicle. Simultaneously, the formation of ice VII is observed, indicating the disproportionation of the $C_2$ phase during its transformation into the $C_3$ phase. The rim around the vesicle predominantly consists of ice VII, with no detectable $C_2$ phase following laser heating. Raman spectra measured at different points in the sample are consistent with this interpretation.

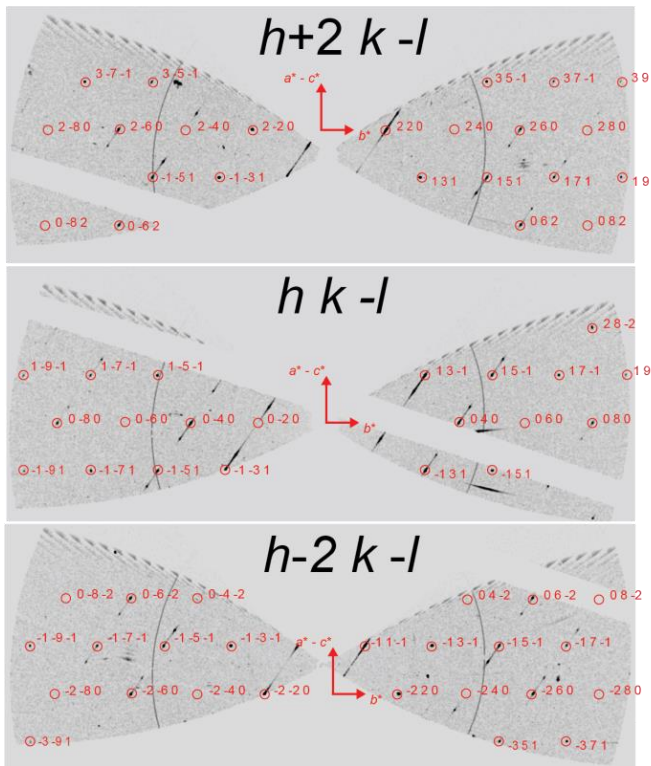

**Figure 3.** Reconstructed reciprocal lattice planes of $Fd\bar{3}m$ C3 phase at 69 GPa. Observed Bragg diffraction spots from sample marked by red circles have been indexed and used to determine the structure (Table 1).

2D diffractograms (Fig. S5 of the Supplemental Materials [12]) reveal that the $C_3$ phase forms as single crystals, evidenced by Bragg peaks appearing as nearly round spots. We applied the SC XRD technique on the highest-quality crystal identified through powder XRD mapping and multigrain analysis (Fig. 3). The structure was solved using ShelXT, a structure solution program that employs the intrinsic phasing algorithm [27-29].

The cubic crystals of the $C_3$-phase grown at 69 GPa exhibited spinel-type twinning, with a mirror plane perpendicular to the body diagonal acting as the twinning operation. Figure 3 presents reciprocal space reconstructions of *hkl* slices for the most intense twin domain, where diffuse streaks along the [$a* + b* - c*$] direction can be identified. These streaks are presumably associated with the presence of twinning. Due to significant reflection overlap (~30%), simultaneous twin integration was applied. However, structure solution and refinement were performed using data collected from the dominant twin component.

An empirical absorption correction was applied using spherical harmonics, implemented within the SCALE3 ABSPACK scaling algorithm, which is included in the CrysAlisPRO software. Following structure solution, the position of the oxygen atom (O1) was identified, while the positions of the hydrogen in the water network (H1) and the hydrogen molecules (H2) were determined from the difference Fourier maps (Table 1). Due to rotational disorder, the hydrogen molecule was approximated as a single hydrogen atom (H2) with an occupancy of 2. No evidence of hydrogen bond symmetrization in the $H_2O$ sublattice was observed at 69 GPa. The H1 atom was split across two positions, with its occupancy fixed at 0.5.

Although a soft constraint was applied to the O-H interatomic distance (0.95 ± 0.05 Å), refinements without this constraint did not significantly affect the position of the H1 atom. The thermal parameters of the hydrogen atoms were freely refined. The crystal structure was refined against $F^2$ using all data by full-matrix least-squares fitting with the SHELXL software [29]. Both the SHELXT and SHELXL programs were implemented within the Olex 2 software package [28]. The final data-to-parameter ratio was ~ 8, while $R_1$ was 4.06%, both indicating the high quality of the refinement.

The data quality was sufficient to determine the complete structure of the $C_3$ phase, including the positions of the $H_2$ molecular centers and the hydrogen atoms in the $H_2O$ network (Table 1). Supplementary crystallographic data for this work have been deposited in the Cambridge Structural Database and are available free of charge from FIZ Karlsruhe (deposition number: CSD2419902) [30].

The crystal structure of the $C_3$ phase at 69 GPa is best described as an $Fd\overline{3}m$ cubic phase, sharing the same space group as the $C_2$ phase. In this structure, oxygen atoms form a face-centered cubic (fcc) lattice, while hydrogen atoms, as part of $H_2O$ molecules, establish hydrogen bonds with adjacent $H_2O$ molecules. This hydrogen-bonded $H_2O$ network remains unchanged in both $C_2$ and $C_3$ phases. The key distinction between the two phases lies in the crystallographic positions occupied by the $H_2$ molecules. Our SCXRD results (Table 1) reveal that these positions are twice as degenerate in the $C_3$ phase compared to the $C_2$ phase (Fig. 4), implying a double capacity for $H_2$ molecules. Interestingly, the $C_3$ phase is observed to be cubic, which contrasts with theoretical predictions [10, 11] that suggest a tetragonal symmetry (space group $P4_1$). We attribute this discrepancy to a three-dimensional hydrogen disorder within the hydrogen-bonded $H_2O$ network. An ordered variant of the $C_3$ phase, in which specific $H_2O$ molecular orientations are realized, would likely emerge at low temperatures. This scenario parallels the well-known relationship between cubic, disordered ice VII and tetragonal, ordered counterpart, ice VIII.

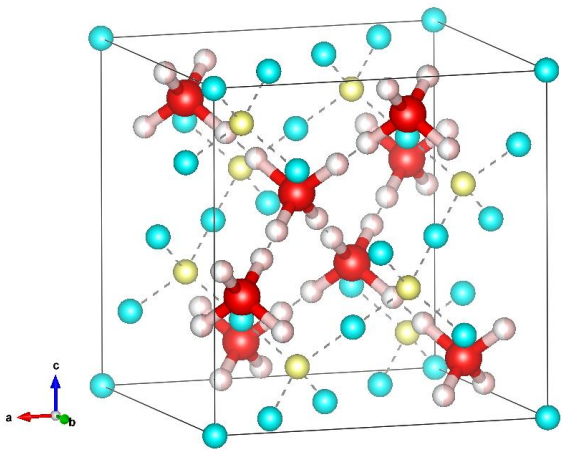

**Figure 4.** Crystal structures of the $C_2$ and $C_3$ phases at 40-69 GPa. Large red spheres represent oxygen atoms, while hydrogen atoms are depicted as smaller spheres in various colors. Light pink spheres denote hydrogen atoms from $H_2O$ molecules, with an occupancy of 0.5, reflecting positional disorder between two adjacent sites. Despite this disorder, the ice rule is maintained, ensuring that each oxygen atom is bonded to exactly two hydrogen atoms. Yellow spheres represent the centers of mass of $H_2$ molecules in the $C_2$ phase, with an occupancy of 2. These $H_2$ molecules are rotationally disordered, and their crystallographic positions are equivalent to those of the oxygen atoms. In the $C_3$ phase, this crystallographic position splits (indicated by dashed lines), and the positions of $H_2$ molecules are shown as cyan spheres.

A significant increase in the $H_2$ molecule composition in the $C_3$ phase compared to the $C_2$ phase is evident from the volume-pressure dependencies shown in Fig. 5. The volume-pressure relationship for the $C_2$ phase aligns closely with results from previous experiments [7, 11]. Our data extend up to 103 GPa, marking the metastability limit of this phase. Notably, the measured volumes are consistently higher than those predicted theoretically for the $Pna2_1$ $C_2$ phase [10], suggesting a higher-than-nominal $H_2$ content in the experimental samples.

In contrast, the experimentally determined volume-pressure relationship for the $C_3$ phase agrees with the theoretical predictions for the $P4_1$ $C_3$ phase [10] presented here and previous experimental observations [11]. Interestingly, the cold-compression data for the $C_3$ phase are systematically lower than those obtained from laser-heated samples but converge at higher pressures. This behavior suggests a gradual uptake of hydrogen during compression in the absence of heating.

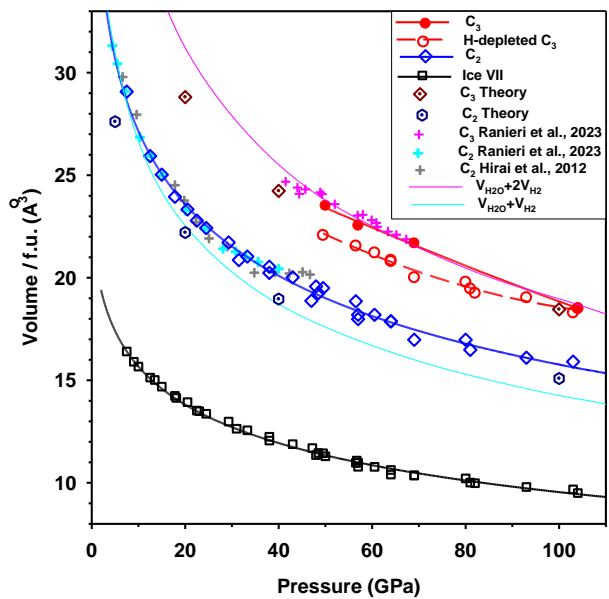

**Figure 5.** Pressure-dependent comparison of the volumes per formula unit for the $C_2$ and $C_3$ compounds, along with ice VII, as determined from experiments and theoretical results in this study and previous work [7, 11]. Lines serve as visual guides. For reference, the combined volumes of ice VII and molecular $H_2$ are also shown, derived from data in this work and Ref. [31].

We applied Raman spectroscopy to probe the hydrogen atomic positions and dynamics within the $C_2$ and $C_3$ phases, with a focus on the structural and vibrational changes occurring during the phase transition. The Raman spectra of the intramolecular stretching modes (vibrons) are particularly sensitive to the local environment of $H_2$ molecules, especially the intra-cage interactions they experience in the $C_2$ and $C_3$ phases. Theoretical calculations predict that the vibron frequency of the $C_3$ phase is lower than in the $C_2$ phase [10]. However, more recent studies [11] suggest a more complex behavior, including the emergence of additional vibron peaks in the $C_3$ phase. Similarly, distinctions in the phonon spectra of the $C_2$ and $C_3$ phases, which involve the translational vibrations of $H_2$ molecules, provide valuable structural insights. For characterizing the hydrogen-bonded $H_2O$ network, the O-H stretching modes are especially informative, as they reflect hydrogen bond strength and may offer evidence for hydrogen bond symmetrization within this subsystem [8, 32].

Raman investigations of the $C_2$-$C_3$ transition were conducted before, after, and occasionally concurrently with the XRD experiments described earlier, across multiple experimental runs. Experiments involving pressure increases at room temperature reveal a continuous evolution in the spectra above 50 GPa (Fig. 6(a)). At low pressures, the high-frequency spectra displayed two prominent peaks: a lower-frequency corresponding to bulk hydrogen and a higher-frequency peak associated with the $C_2$ phase.

Above 50 GPa, the vibron mode of the $C_2$ phase broadened, and a new higher-frequency vibron peak emerged. Concurrently, significant changes were observed in the low-frequency region of the spectra during compression (Fig. S6 of the Supplemental Materials [12]). The quantum rotational (roton) spectra of $H_2$ molecules [33] evolved continuously with increasing pressure. The $S_0(1)$ peak, initially located near 600 cm$^{-1}$, split and transitioned into a triplet at 52 GPa, accompanied by the emergence of a weaker peak near 260 cm$^{-1}$ (Fig. S6 of the Supplemental Materials [12]).

At 67 GPa, an additional vibron peak appeared between the original and the newly formed peaks, with all these new peaks gaining intensity as pressure increased. Between 80 and 103 GPa, after several weeks, the spectra evolved to a broad, asymmetric main vibron peak with a weak, higher-frequency shoulder persisting (Fig. 6(a)). Upon decompression, the spectral features remained similar but shifted to lower frequencies. At 53 GPa, the vibron peak became slightly narrower, although its asymmetry persisted.

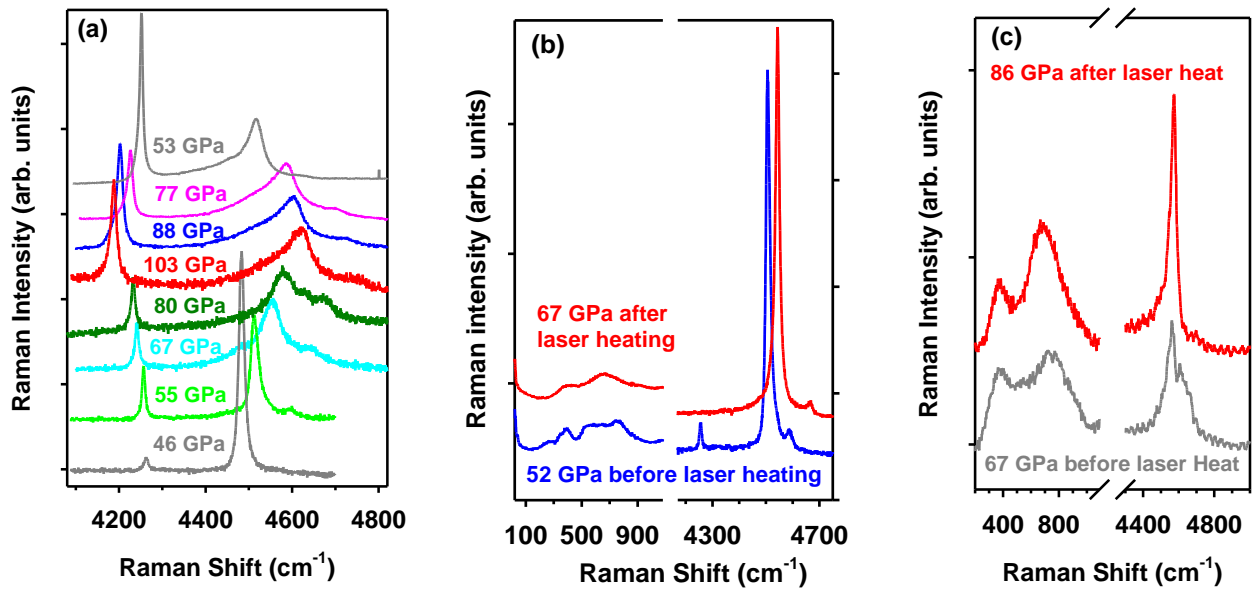

**Figure 6.** Raman spectra at the $C_2$–$C_3$ transition. (a) Continuous transformation of the spectra during compression from 46 GPa to 103 GPa, followed by decompression to 53 GPa. (b) and (c) Spectral changes induced by laser heating at 52 GPa and 67 GPa, respectively. The excitation wavelengths used are 532 nm, 488 nm, and 630 nm.

In laser-heating experiments conducted at 52 GPa and 67 GPa, the Raman spectra of the initial state exhibit additional roton and vibron peaks, consistent with those observed in the cold-compression experiments. Following laser heating, the spectra undergo modifications in both the rotational and vibrational regions (Fig. 5(b, c) and Fig. S6 of the Supplemental Materials [12]). Specifically, the additional Raman peak within the rotational mode spectral range disappears, while the vibron spectra undergo subtle changes. These changes include a frequency decrease of approximately 10 cm$^{-1}$ and the emergence of a weak shoulder on the lower-frequency side of the main vibron peak, or a transformation of the peak shape into an asymmetric profile (Fig. 6(b,c) and Fig. S7 of the Supplemental Materials [12]).

At higher pressures, laser heating leads to a narrowing of the vibron bands and the appearance of a weak additional high frequency vibron band (Fig. 6(c)). In contrast, no visible changes in the Raman spectra were observed after laser heating at 32 GPa.

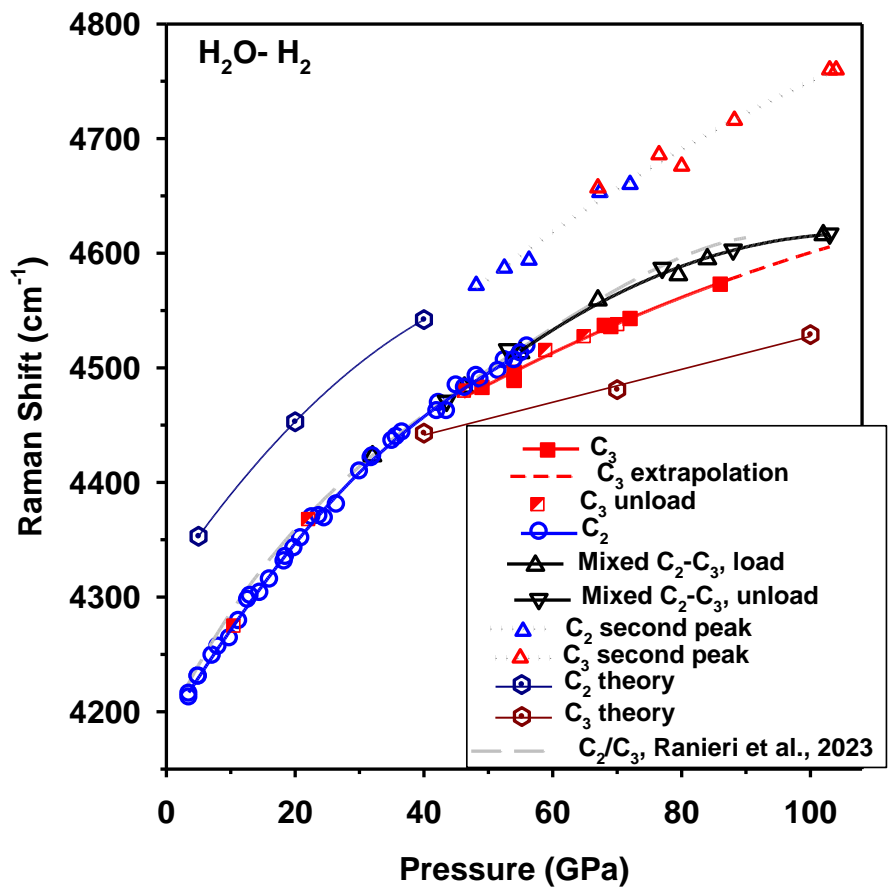

**Figure 7.** Pressure dependence of the Raman vibron frequencies for the $C_2$ and $C_3$ phases, as determined from this work (experiment and theory) and from previous experimental studies [11].

The pressure dependence of the main vibron Raman band of the $C_2$ phase exhibits a sublinear trend (Fig. 7). Above 50 GPa, cold compression results in the coexistence of $C_2$ and $C_3$ phases (Fig. 1). In this regime, the pressure evolution of the main vibron peak continues from the lower-pressure $C_2$ curve and shows a turnover near 100 GPa. Upon laser heating in this pressure range, the sample transitions fully to the $C_3$ phase, which is accompanied by a softening of the main vibron band. When extrapolated to higher pressures, the pressure dependence of the main vibron frequency nearly converges with that of the cold-compressed sample (Fig. 7). This observation is consistent with XRD results (Fig. 1) indicating that the cold-compressed sample at 104 GPa is in the $C_3$ phase.

Interestingly, no hysteresis is observed in the pressure dependence of the main vibron frequency during unloading for the cold-compressed sample. In contrast, the laser-heated sample—after exhibiting a frequency drop during the $C_2$-$C_3$ phase— continues to follow a pressure dependence of the vibron frequency corresponding to the $C_3$ phase. This trend persists until the $C_3$ phase vibron frequency converges with the $C_2$ vibron frequency-pressure dependence at approximately 45 GPa. The second, weaker vibron peak, observed above 47 GPa, increases nearly linearly with pressure (Fig. 7). Notably, no significant differences in its frequency were detected among the $C_2$, $C_3$, and mixed states.

Theoretical calculations addressing the phase transition between the $C_2$ and $C_3$ phases were conducted by approximating the real crystal structures of these phases using the hydrogen ordered $Pna2_1$ and $P4_1$ structures, respectively, as described above [10]. The $P4_12_12$ variant of the $C_2$ phase was found slightly less energetically favorable than $Pna2_1$, consistent with Ref. [10]. Both the $C_2$ and $C_3$ phases feature a sublattice of hydrogen-bonded $H_2O$ molecules. At 40 GPa, the $Pna2_1$ $C_2$ approaches the regime of symmetric hydrogen bonding, whereas in the $P4_1$ $C_3$ phase, the difference in the intra- and interatomic O-H lengths is substantial at the same pressure (1.51 vs 1.01 Å). However, by 100 GPa, the $P4_1$ $C_3$ phase also develops symmetric hydrogen bonds.

As shown by this and previous investigations, the real structures of these phases are cubic (or nearly cubic) within the relevant pressure range, where they are stable or near thermodynamic stability. Our theoretical calculations are consistent with prior studies [10, 11] regarding the stability limits of the $C_2$ and $C_3$ phases. We find that the $C_2$ phase remains stable up to 14 GPa, while the $C_3$ phase becomes stable above 23 GPa. Both phases are metastable in the intermediate pressure range between 14 and 23 GPa (Fig. S8 of the Supplemental Materials [12]).

The theoretically calculated Raman spectra for the $C_2$ and $C_3$ compounds exhibit significant activity across three characteristic spectral regions (Fig. 8). In these calculations, the $C_2$ and $C_3$ compounds were modeled using the hydrogen-ordered $Pna2_1$ and $P4_1$ structures, respectively. The results of the group theory analysis of the Raman-active vibrational modes for these structures are provided in Table S2 of the Supplemental Materials [12].

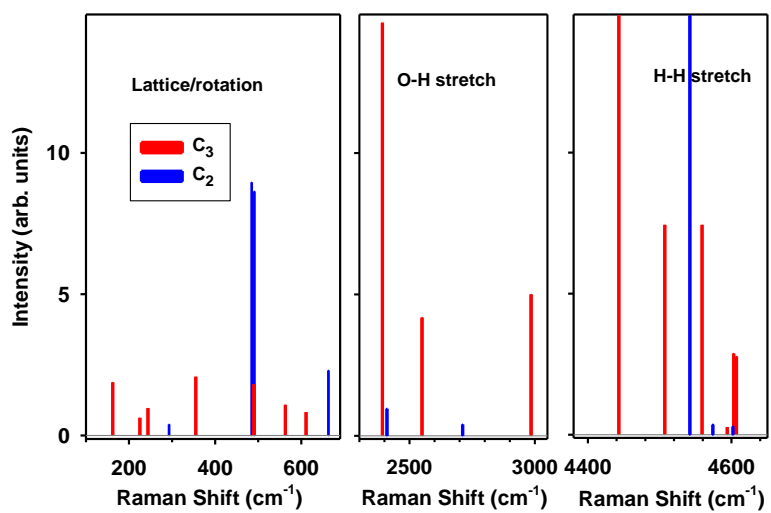

**Figure 8.** Theoretical Raman spectra calculated at 40 GPa for the $C_2$ and $C_3$ compounds, modeled using hydrogen-ordered $Pna2_1$ and $P4_1$ structures, respectively. The main vibron modes in the high-frequency region (one for each compound) are approximately ten times more intense than the other modes and are shown with intensity saturation to enhance the visibility of weaker spectral features. Direct comparison with experimental data (Fig. 6 and Figs. S6, S7 of the Supplemental Materials [12]) is limited due to the quantum nature of the rotational modes of hydrogen molecules and the overdamped character of the O-H stretching modes near the regime of hydrogen bond symmetrization.

In the low-frequency range, up to approximately 700 cm$^{-1}$, the Raman spectra display multiple modes associated with molecular rotations of $H_2O$ and $H_2$ molecules, as well as their lattice vibrations (Fig. S10 of the Supplemental Materials [12]). The H-O-H bending modes of the $H_2O$ molecules are predicted to appear in the 1700-1850 cm$^{-1}$ range, although these modes are typically too weak to be observed. The O-H stretching modes are predicted in the 2400-3100 cm$^{-1}$ frequency region, with their frequencies depending on the strength of the hydrogen bonding between $H_2O$ molecules. In the regime of hydrogen bond symmetrization, the frequency of the symmetric O-H stretching mode softens below 1000 cm$^{-1}$ and interacts with other modes of the same symmetry in this spectral range [32]. Our calculations for the $C_2$ compound at 40 GPa are consistent with these predictions (Fig. S10 of the Supplemental Materials [12]). Notably, the eigenvector of the 663 cm$^{-1}$ $A_1$ mode includes a substantial contribution from the O-H stretching vibrations.

The internal $H_2$ molecular vibrations are represented by the H-H stretch modes (vibrons), observed in the 4400–4700 cm$^{-1}$ spectral range. During the $C_2$ to $C_3$ transition, the Raman activity of these modes is expected to increase due to the doubling of the number of $H_2$ molecules in the $C_3$ phase (Fig. 9, Table S2 of the Supplemental Materials [12]). In both phases, the Raman spectra of the vibrons are dominated by a single fully symmetric band, in which all $H_2$ molecules vibrate in phase (the main vibron mode). Additional vibron modes (sidebands) correspond to various combinations of $H_2$ molecule vibrations and exhibit *higher* frequencies than the main $H_2$ vibron mode (cf. Ref. [11]), consistent with similar observations in compressed bulk hydrogen [35]. As in bulk hydrogen [36], these sidebands correspond to different points in the extended vibron band structure, with the total vibron frequency spread reflecting the vibron bandwidth, which is determined by intermolecular coupling. Upon the $C_2$-to-$C_3$ transition, the main vibron bands shifts to lower frequencies (Figs. 7, 9), as previously predicted in Ref. [10]. The vibron bandwidth increases in the $C_3$ phase compared to the $C_2$ phase due to stronger intermolecular coupling, resulting from the closer packing of $H_2$ molecules in the $C_3$ phase.

Our results establish the structure and composition of the $C_3$ phase and shed light on the mechanism underlying the $C_2$ to $C_3$ transformation. We find that the $C_2$ to $C_3$ transition occurs gradually at 300 K at elevated pressures above 50 GPa, accelerated by the presence of bulk $H_2$ reservoir within the high-pressure cavity. Notably, the transition also proceeds under near-stoichiometric conditions, with an approximate 1:1 $H_2$:$H_2O$ ratio, as it is in the $C_2$ phase, albeit at high temperatures. Ice VII appears in the high-pressure cavity as a result of hydrogenation of the $C_2$ phase, whether induced by pressure or temperature. Previous investigations reported the onset of the $C_2$ to $C_3$ transition at 40–45 GPa upon laser heating to 1200(200) K [11]. Our results confirm that this transition pathway remains viable across a broader pressure range from 47 to 86 GPa.

Theoretical analysis of the pressure-dependent energetics of the $C_2$ and $C_3$ phases supports previous conjectures regarding their metastability between 14 and 23 GPa. This conclusion aligns with earlier observations [11, 37] and findings from this study, where the Raman band of bulk hydrogen appears above 20 GPa (Fig. S6 of the Supplemental Materials [12]) due to partial decomposition of the $C_2$ compound. In this study, this behavior was evident in $H_2$-rich samples, as shown by the strong presence of the bulk hydrogen vibron. Above 23 GPa, the $C_3$ phase becomes thermodynamically stable, with its stability increasing with pressures. At 100 GPa, the formation energy of $C_3$ exceeds 15 meV (Fig. S8 of the Supplemental Materials [12]), enabling spontaneous transition at room temperature without requiring additional thermal activation. In contrast, the $C_2$ phase is unstable at this pressure, with a formation energy of -111 meV.

Based on the results of SC XRD measurements, we find that the structure of the $C_3$ phase is cubic, in contrast to the lowest-enthalpy phases predicted by theory [10]. Earlier experimental studies, relying on powder XRD patterns, were limited to observing only two major Bragg reflections which restricted the ability to directly resolve the hydrogen composition and crystallographic positions. In contrast, our SC XRD measurements detected Bragg reflections from 41 different *hkl* indices, allowing precise determination of the hydrogen-bond network and the positions of $H_2$ molecules within the unit cell (Table 1). Structural refinement at 69 GPa required considerations of twinning, indicating that the observed cubic symmetry likely represents an average over nanoscale domains that may be less symmetric. In addition, there are indications (diffuse streaks) of stacking faults or lattice disorder (Fig. 3), which require a separate investigation. The experimentally observed cubic unit cell is twice the volume of the theoretically predicted tetragonal one, with the lattice parameters of the tetragonal ($a_t$) and cubic ($a_c$) lattices related by $a_c = \sqrt{2}\, a_t$. Our theoretical calculations indicate that the deviation from cubic symmetry is less than 1% as indicated by the $(\sqrt{2}a_t)/c_t$ ratio (see also Ref. [11]). This small deviation justifies the use of the tetragonal $P4_1$ structure in theoretical modeling and reinforces its relevance to interpreting the experimental results.

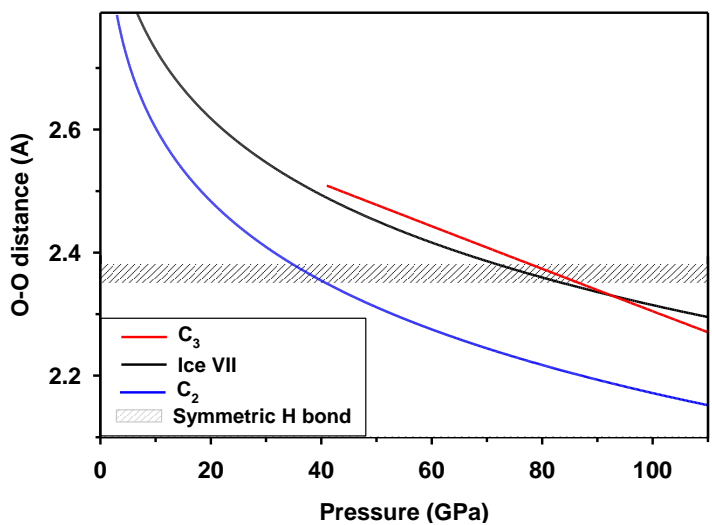

**Figure 9.** Experimentally determined O-O distances in the $C_2$ and $C_3$ phases as a function of pressure. The hatched region indicates the O-O distance range where the hydrogen bond symmetrization occurs [32, 38].

The structural data obtained in this study provide insights into hydrogen bond symmetrization in the $C_2$ and $C_3$ phases. We analyzed the volume versus pressure data for the $C_2$, $C_3$, and ice VII phases to determine the shortest O-O distances (Fig. 9). Hydrogen bond symmetrization occurs at pressures of 32–40 GPa in the $C_2$ phase and 78–87 GPa in the $C_3$ phase, consistent with previous observations for the $C_2$ phase [8]. These results demonstrate that hydrogen bond symmetrization in the $C_3$ phase is significantly pressure-delayed compared to the $C_2$ phase, despite the structural similarities in their $H_2O$ networks. The higher symmetrization pressure in the $C_3$ phase likely reflects the influence of increased $H_2$ content and stronger intermolecular interactions, which stabilize asymmetric hydrogen bonds to higher pressures.

Our theoretical calculations indicate nearly symmetric hydrogen bonds in the $C_2$ phase at 40 GPa (1.11 and 1.21 Å) and fully symmetric bonds in the $C_3$ phase at 100 GPa (1.15 Å). Hydrogen bond symmetrization in the $C_3$ phase is expected to occur near the pressure observed in pure $H_2O$, approximately 75 GPa, based on the spectroscopic data of Ref. [32]. However, this interpretation remains a subject of debate [39, 40]. The main challenge lies in accurately accounting for proton dynamics in the regime near symmetrization, where the proton is believed to tunnel between two potential wells corresponding to different O-H bond geometries.

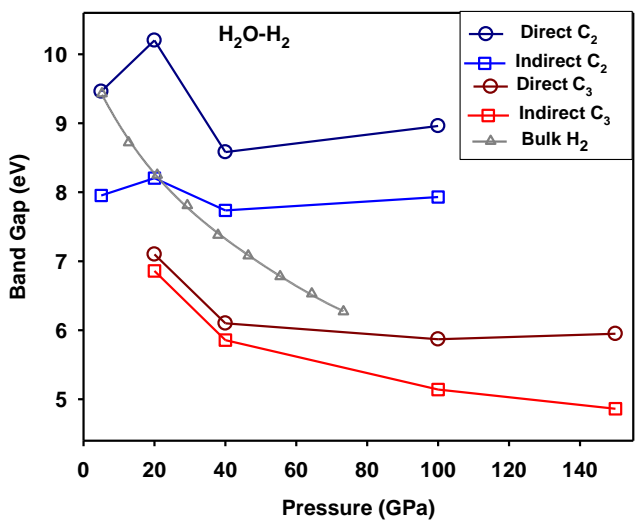

**Figure 10.** Theoretically calculated band gap values for the $C_2$ and $C_3$ compounds as a function of pressure. These results are compared with theoretical predictions for the band gap of bulk hydrogen, which are in good agreement with the most reliable experimental determinations [34].

Electronic structure calculations show that the $C_2$ and $C_3$ phases are insulating (Fig. 10 and Fig. S9 of the Supplemental Materials [12]), with the smallest bandgap being indirect in both compounds. In the $C_2$ phase, the bandgap exhibits weak pressure dependence, whereas in the $C_3$ phase, it decreases much more rapidly with pressure (Fig. 10). This difference stems from the highly pressure-dependent electronic band of $H_2$ molecules at the center of the Brillouin zone. At high pressures, these molecules are in close proximity (approximately 1.94 Å between molecular centers at 100 GPa), which is comparable to or even shorter than the intermolecular distance in bulk hydrogen phase I at the same pressure. However, above 40 GPa, the bandgap of the $C_3$ phase begins to decrease more gradually, indicating that $H_2$ cavities become less compressible in this regime. This behavior is likely due to the constraining effect of the surrounding $H_2O$ network, which limits further lattice contraction.

Our Raman spectroscopy data, supported by theoretical calculations, offer new insights into the $C_2$ to $C_3$ phase transition, as well as the structural and hydrogen compositional characteristics of these phases. The most significant findings concern the behavior of the main Raman vibron mode in the $C_2$ and $C_3$ compounds, which corresponds to the in-phase stretching vibrations of hydrogen molecules. Notably, the frequency of this mode is higher in both compounds compared to bulk $H_2$. This difference arises because, in bulk hydrogen, the vibron frequency is significantly lowered by strong intermolecular coupling, which leads to a downward shift [35].

In inclusion compounds such as $C_2$ and $C_3$, the $H_2$-$H_2$ intermolecular coupling is reduced due to the lower number of nearest-neighbor $H_2$ molecules. Consequently, the vibron frequency in the hydrogen-rich $C_3$ compound is expected to be smaller than in $C_2$. This trend is consistent with theoretical calculations presented in Ref. [10] and this work (Fig. 7) and contrasts with the conjectures of Ref. [11]). Our experimental results show the anticipated redshift in vibron frequency associated with the $C_2$ to $C_3$ phase transition (Fig. 7). Although this shift is smaller than predicted theoretically, it remains significant and measurable.

Above 47 GPa during cold compression, the Raman spectra of the $C_2$ phase undergo changes (Fig. 6, Fig. S6 of the Supplemental Materials [12]), likely due to the gradual diffusion of hydrogen into the cages within the $H_2O$ network structure. This is consistent with an increased volume of the $C_2$ phase compared to calculations (Fig. 5) as well as the broadening and splitting of the diffraction peaks (Fig. 1), which occurs concomitantly with the appearance of grains of the $C_3$ phase as pressure increases. In this regime, the Raman spectra reveal additional peaks in the region associated with translational and rotational $H_2$ vibrations (near 550 cm$^{-1}$) and several peaks in the high-frequency range above the position of the main $C_2$ hydrogen vibron.

Theoretical calculations predict additional Raman activity in the $C_3$ phase (Fig. 9); however, the intensity of all modes other than the main Raman vibron is expected to be low. We speculate that the observed spectral features in the mixed $C_2$-$C_3$ regime arise from local lattice distortions and violations of Raman selection rules. These factors likely cause the splitting of the $S_0(0)$ and $S_0(1)$ roton peaks as well as the appearance of additional vibron sidebands.

In the $C_3$ phase synthesized via cold compression at 104 GPa, the Raman spectra of the main vibron mode remains broad and asymmetric even after the transformation is complete (Fig. 6(a)), suggesting that disorder in the positioning of $H_2$ molecules persists. In contrast, the $C_3$ phase synthesized via laser heating exhibits a narrow vibron mode with no additional low-frequency peaks. A weak higher- frequency vibron sideband is observed in this case, in agreement with theoretical predictions (Fig. 6(b), Fig. 8).

It is important to stress that our theoretical calculations predict weaker Raman vibron modes at higher frequencies compared to the main vibron mode. These modes involve collected molecular vibrations of all $H_2$ molecules within the unit cell. This interpretation contrasts with that of Ref. [11], where the weaker vibron modes were attributed to local vibrations arising from different molecular environments. In our view, supported by theoretical calculations, all $H_2$ molecules coupled via intermolecular interactions contribute uniformly to the vibron optical modes, as they occupy equivalent crystallographic sites with the same symmetry ($16c$ or $C_{3i}$). Variations in amplitude and phase among these modes (affecting their relative intensities) arise from differences in their vibrational symmetries.

Interestingly, secondary Raman vibron modes were reported at lower frequencies than the main vibron mode in the experiments described in Ref. [11]. Based on the broadness of these bands and the main vibron band, we speculate that the samples in Ref. [11] were likely disordered and stressed, similar to those obtained in our experiments via cold compression. Heating the samples to higher temperatures would likely anneal them, reducing stress gradients and mitigating structural and compositional nonuniformities.

## IV. CONCLUSIONS

Our combined experimental and theoretical investigations of the $C_2$ to $C_3$ phase transition, as well the structure and properties of these phases, have provided several significant insights into their behavior and underlying mechanisms:

1. We have demonstrated that the $C_2$ to $C_3$ transition can be achieved through cold compression under $H_2$-rich conditions at pressures as high as 103 GPa. This transition occurs gradually and is marked by both structural and vibrational changes, offering a clearer understanding of the pressure-induced phase behavior of these hydrogen-water inclusion compounds.

2. Single-crystal XRD analysis reveals that the $C_3$ phase crystallizes in the cubic space group $Fd\bar{3}m$, which is isostructural with the $C_2$ phase with respect to the positions of the oxygen atoms. The key difference lies in the arrangement of hydrogen molecules: in the $C_3$ phase, $H_2$ molecules occupy the 16c positions, whereas in the $C_2$ phase, they are positioned in the 8b sites. This structural distinction indicates a doubling of the $H_2$ molecule density in the $C_3$ phase compared to the $C_2$ phase, offering a crucial structural insight into densification mechanism and the nature of the $C_2$ to $C_3$ phase transition.

3. Raman spectroscopy across the $C_2$ to $C_3$ transition reveals a notable downward shift in the main vibron mode frequency. This frequency shift supports the structural changes identified via XRD, particularly the increased $H_2$ density in the $C_3$ phase. The observed broadening and changes in the spectral features confirm intensified molecular interaction during the transformation, as the system transitions into the more hydrogen-rich $C_3$ phase.

4. Our theoretical calculations of the electronic structure of the $C_2$ and $C_3$ phases under high pressure reveal a pressure induced narrowing of the bandgap of the $C_3$ phase primarily driven by enhanced intermolecular coupling between $H_2$ molecules. However, this effect appears to level off at pressures near 100 GPa, indicating that the system reaches a saturation point in intermolecular interactions. At this stage, further compression yields only minor changes in the electronic properties, suggesting that the structural framework of the $H_2O$ network imposes a limiting constraint on the compressibility of $H_2$ cavities.

## ACKNOWLEDGEMENT

Support is acknowledged from the National Science Foundation Grant No. CHE-2302437 and Carnegie Science. E.B. and M.B. acknowledge the support from the DFG Emmy-Noether Program (Projects BY101/2-1 and BY112/2-1) and by the Johanna-Quandt Stiftung. M.B. acknowledges the support by the LOEWE program. Portions of this work were performed at GeoSoilEnviroCARS (The University of Chicago, Sector 13), Advanced Photon Source (APS), Argonne National Laboratory. GeoSoilEnviroCARS is supported by the National Science Foundation – Earth Sciences (EAR – 1634415). This research used resources of the Advanced Photon Source, a U.S. Department of Energy (DOE) Office of Science User Facility operated for the DOE Office of Science by Argonne National Laboratory under Contract No. DE-AC02-06CH11357. We acknowledge DESY (Hamburg, Germany), a member of the Helmholtz Association HGF, for the provision of experimental facilities. Parts of this research were carried out at PETRA III (beamline P02.2). We thank A. Steele for help in Raman mapping at Carnegie.

**Table 1. Details of crystal structure refinement for $Fd\bar{3}m$ $H_2O(H_2)_2$ at 69 GPa**

| | |
|---|---|
| Empirical formula | $H_2O(H_2)_2$ |
| Formula weight, Mr | 22.05 |
| Crystal system | Cubic |
| Space group | $Fd\bar{3}m$ |
| a/Å | 5.5773(4) |
| b/Å | 5.5773(4) |
| c/Å | 5.5773(4) |
| $\alpha/°$ | 90 |
| $\beta/°$ | 90 |
| $\gamma/°$ | 90 |
| Volume/Å$^3$ | 173.49(4) |
| Z | 8 |
| ρcalc g/cm$^3$ | 1.688 |
| μ /mm$^{-1}$ | 0.045 |
| F(000) | 112.0 |
| Crystal size /mm$^3$ | $0.01 \times 0.01 \times 0.01$ |
| Radiation | synchrotron ($\lambda = 0.28457$) |
| 2Θ range for data collection/° | 5.066 to 35.008 |
| Index ranges | $-9 \leq h \leq 9$, $-8 \leq k \leq 9$, $-3 \leq l \leq 5$ |
| Reflections collected | 124 |
| Independent reflections | 41 [Rint = 0.0576, Rsigma = 0.0367] |
| Data/restraints/parameters | 41/1/5 |
| Goodness-of-fit on F2 | 1.211 |
| Final R indexes [I>=2σ (I)] | R1 = 0.0406, wR2 = 0.0875 |
| Final R indexes [all data] | R1 = 0.0439, wR2 = 0.0896 |
| Largest diff. peak/hole / e Å$^{-3}$ | 0.26/-0.28 |

| Atomic positions | Site symmetry | x | y | z | $U_{iso}*/U_{eq}$ | occupancy |
|---|---|---|---|---|---|---|
| O1 | 8b | 0.125 | 0.125 | 0.625 | 0.0071(3) | 1 |
| H1 | 32e | 0.034(4) | 0.034(4) | 0.534(4) | 0.018(11)* | 0.5 |
| H2 | 16c | 0 | 0 | 0 | 0.051(7)* | 2 |

# Supplementary Information

# Hydrogen-rich hydrate at high pressures up to 104 GPa

Alexander F. Goncharov[1], Elena Bykova[2], Iskander Batyrev [3], Maxim Bykov[4], Eric Edmund[1], Amol Karandikar[1], Mahmood Mohammad[5], Stella Chariton[6], Vitali Prakapenka[6], Konstantin Glazyrin[7], Mohamed Mezouar[8], Gaston Garbarino[8], Jonathan Wright[8]

[1] Earth and Planets Laboratory, Carnegie Science, Washington, DC 20015, USA
[2] Goethe-Universität Frankfurt am Main, Facheinheit Mineralogie, 60438 Frankfurt am Main, Germany
[3] U.S. Army DEVCOM Army Research Laboratory, FCDD-RLA-WA, Aberdeen Proving Ground, Maryland, 21005 U.S.A.
[4] Institute of Inorganic and Analytical Chemistry, Goethe University Frankfurt, Max-von-Laue-Straße 7, 60438 Frankfurt am Main, Germany
[5] Department of Mathematics, Howard University, Washington DC 20059 USA.
[6] Center for Advanced Radiation Sources, The University of Chicago, Chicago, Illinois 60637, USA
[7] Deutsches Elektronen-Synchrotron DESY, Notkestr. 85, 22607 Hamburg
[8] European Synchrotron Radiation Facility BP 220, 38043 Grenoble Cedex, France

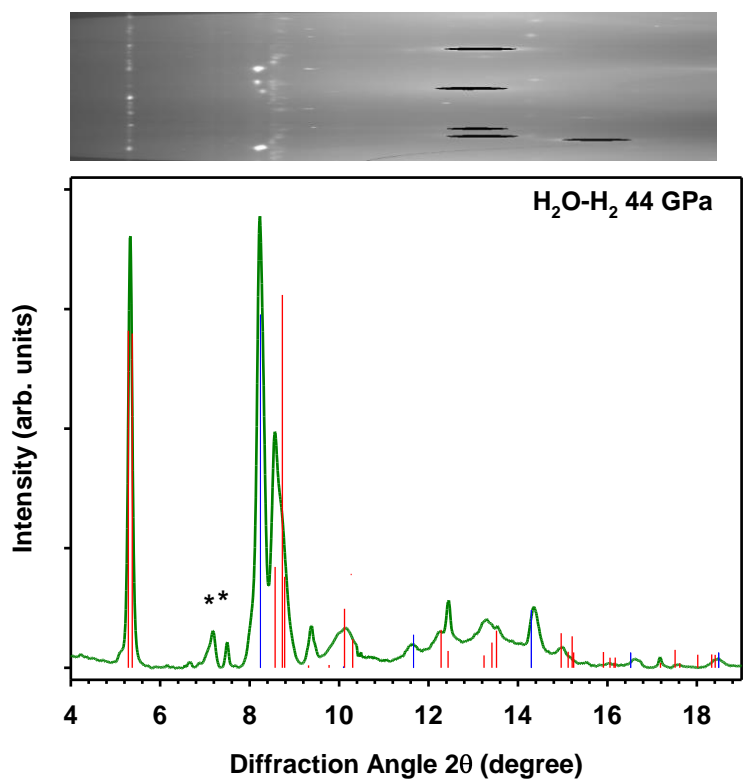

**Figure S1.** X-ray diffraction patterns measured at 44 GPa at 300 K at Petra III, P02.2. The top panel displays 2D diffractograms in rectangular coordinates. Vertical lines show the calculated positions of diffraction lines of $Pna2_1$ $C_2$ at 44 GPa. Asterisks mark Bragg peaks originating from the rhenium gasket. The X-ray wavelength used is 0.2919 Å.

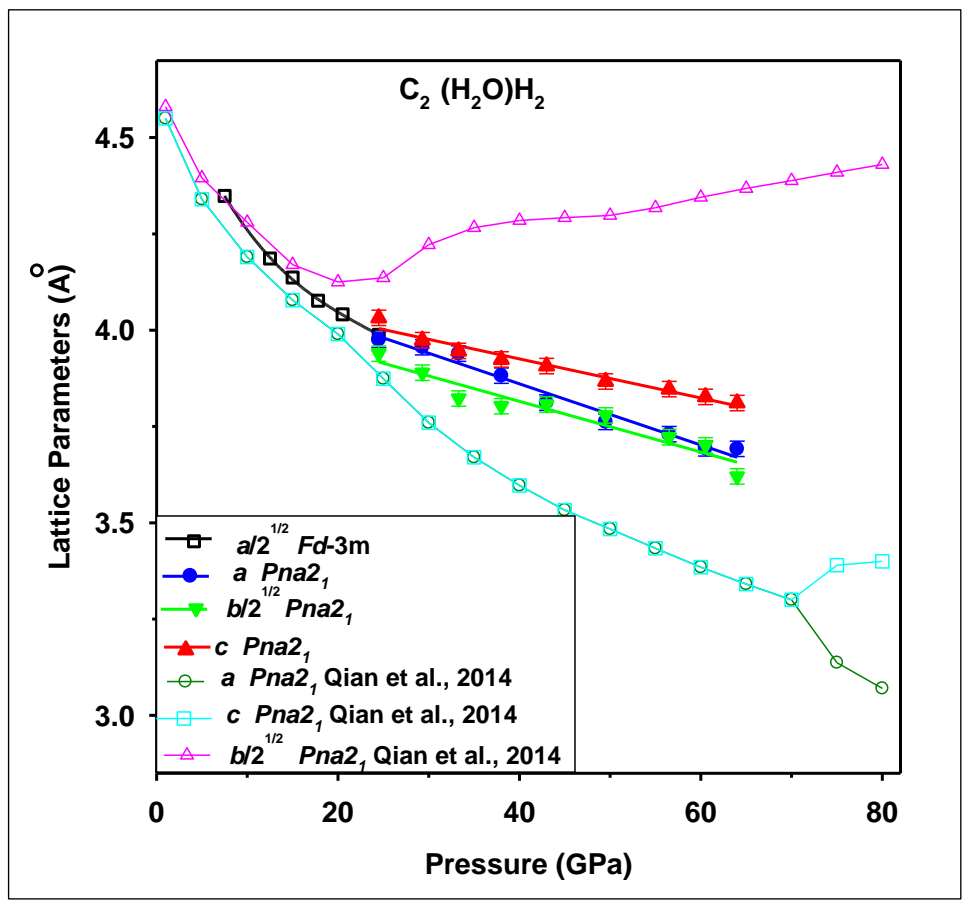

**Figure S2.** Lattice parameters of the C₂ H₂O-H₂ assuming the *Pna*2₁ structure predicted theoretically. The data are scaled as indicated in the legend for detailed comparison, highlighting the quasicubic symmetry of this phase.

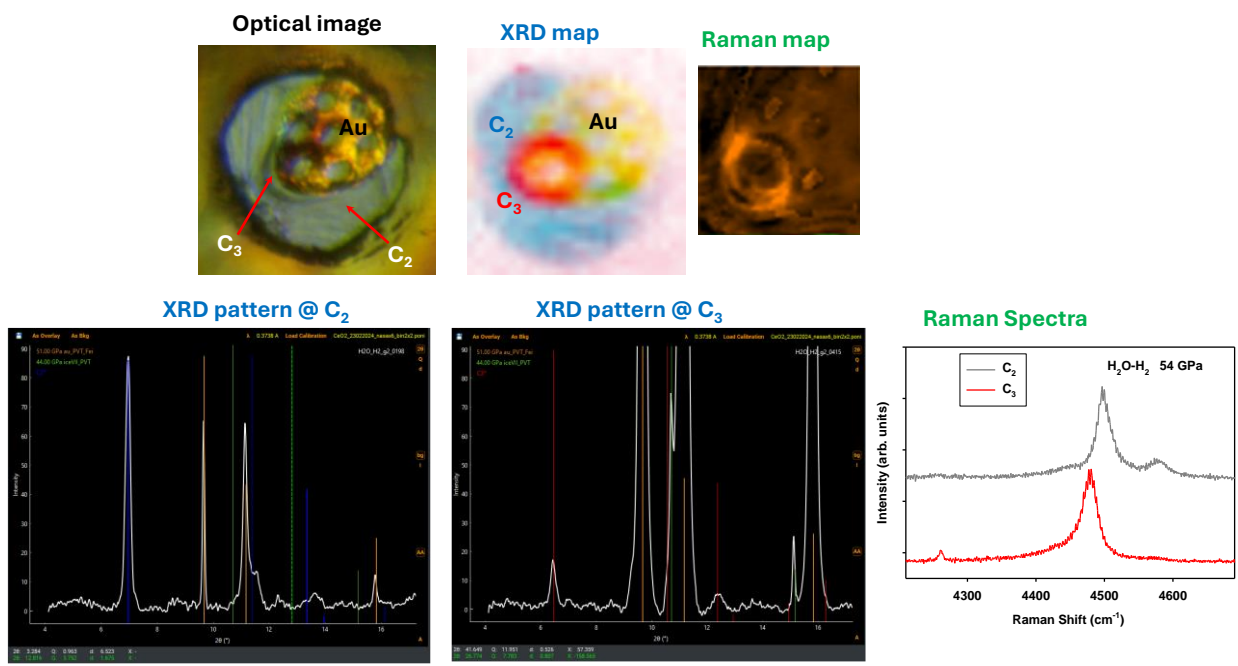

**Figure S3.** Formation of the $H_2O$-$H_2$ $C_3$ compound after near-IR laser heating via Au coupler. The top row presents optical, XRD, and Raman images, while the bottom row shows the 1D integrated XRD patterns at the positions of the $C_2$ and $C_3$ compounds. Additionally, the Raman spectra are displayed for two sample positions corresponding to the $C_2$ and $C_3$ compounds.

**Optical image**

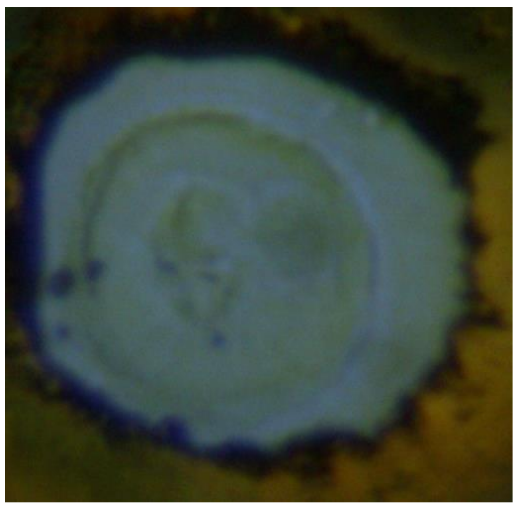

**XRD map**

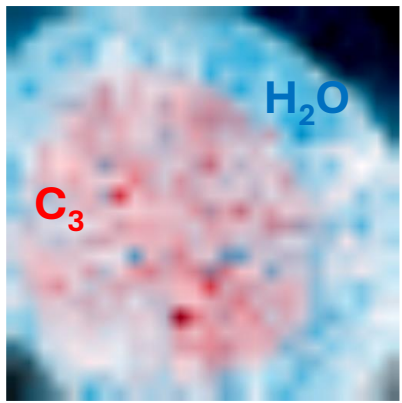

**Figure S4.** Formation of the $H_2O$-$H_2$ $C_3$ compound after IR $CO_2$ laser heating at 52 GPa; pressure increased to 69 GPa after laser heating. The left image depicts the emergence of a vesicle in the center of the high-pressure cavity. XRD mapping reveals that the vesicle consists of a mixture of the $C_3$ compound and ice VII, while the surrounding region is predominantly ice VII (Fig. S5). This conclusion is also corroborated by Raman spectroscopy observations as detailed in the main text.

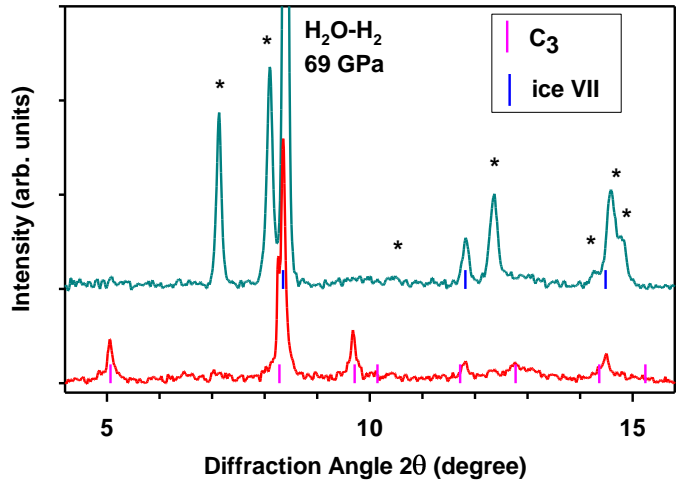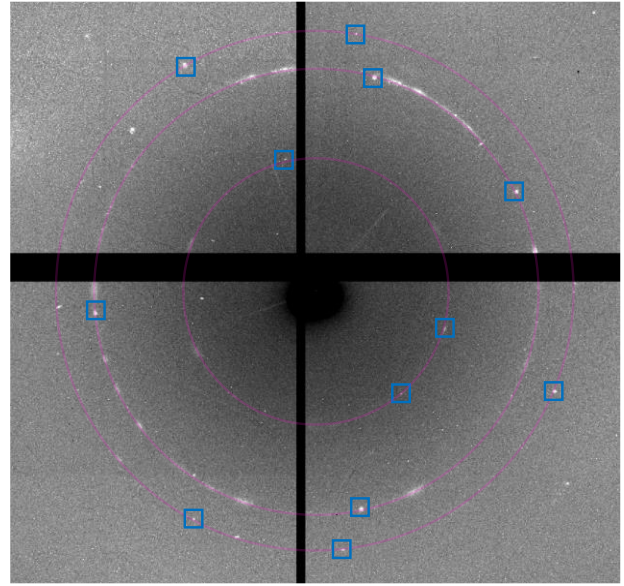

**Figure S5.** X-ray diffraction patterns measured at 69 GPa and 300 K at ESRF, Sector 11 after IR $CO_2$ laser heating at 52 GPa (left panel). The vertical lines indicate the calculated positions of diffraction lines of the $C_3$ filled ice and ice VII phases. Asterisks mark Bragg peaks originated from the rhenium gasket. The right panel displays a 2D diffractogram in the $C_3$ phase with Bragg reflections marked by blue squares. The calculated positions of diffraction lines of the $C_3$ filled ice are shown by semitransparent pink circles. The X-ray wavelength used is 0.2846 Å.

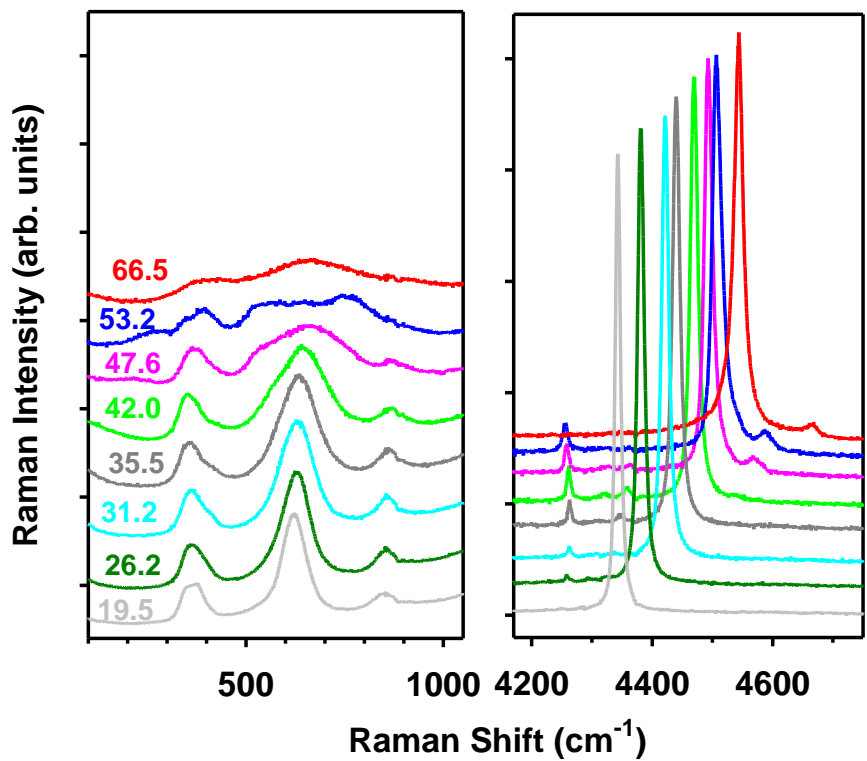

**Figure S6.** Raman spectra during compression from 19.5 GPa to 53.2 GPa at 300 K, followed by $CO_2$ laser heating, which resulted in a pressure increase to 66.5 GPa. The excitation wavelength used is 488 nm. The data illustrate the evolution of Raman-active modes as pressure increases, highlighting changes induced by compression and subsequent laser heating.

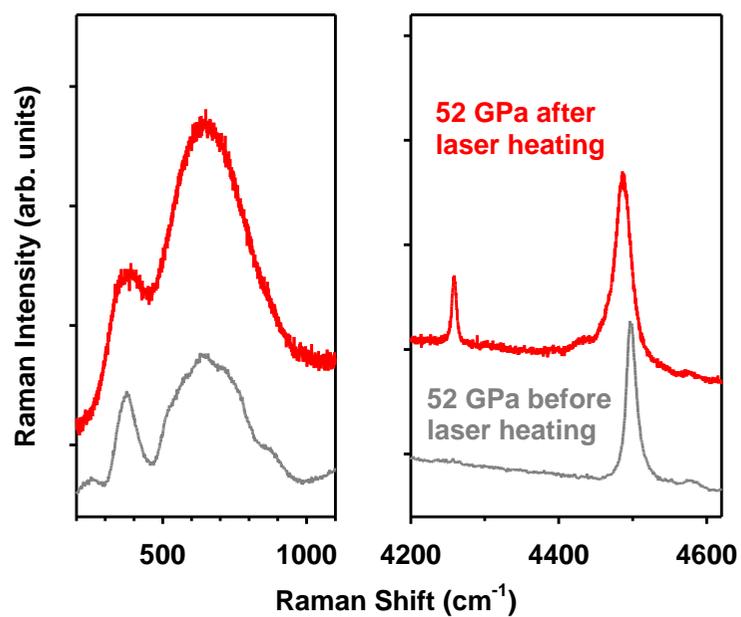

**Figure S7.** Raman spectra at the $C_2$–$C_3$ transition induced by laser heating at 52 GPa. The excitation wavelength used is 660 nm.

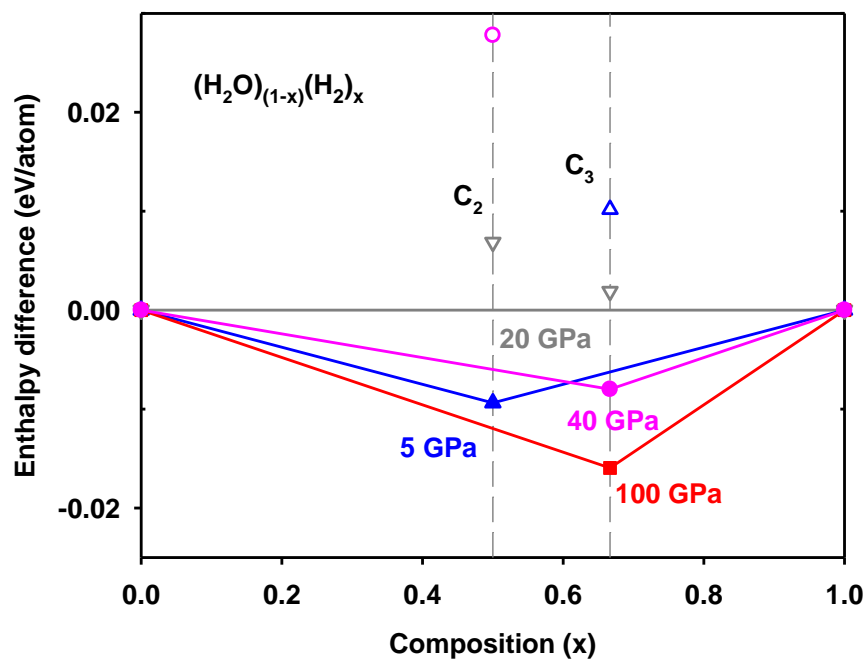

**Figure S8.** Convex Hull diagram for the $(H_2O)_{(1-x)}(H_2)_2$ system at selected pressures. This figure illustrates the enthalpy of formation of the $C_2$ and $C_3$ compounds derived from the $H_2O$ and $H_2$ end member materials. The structures of $H_2O$ and $H_2$ are represented as $I4_1/amd$ ice VIII and $P6_3/m$ ordered $H_2$-I, respectively. The $C_2$ and $C_3$ compounds are modeled by hydrogen ordered $Pna2_1$ and $P4_1$ structures. Filled symbols indicate stable compounds, while open symbols represent metastable compounds.

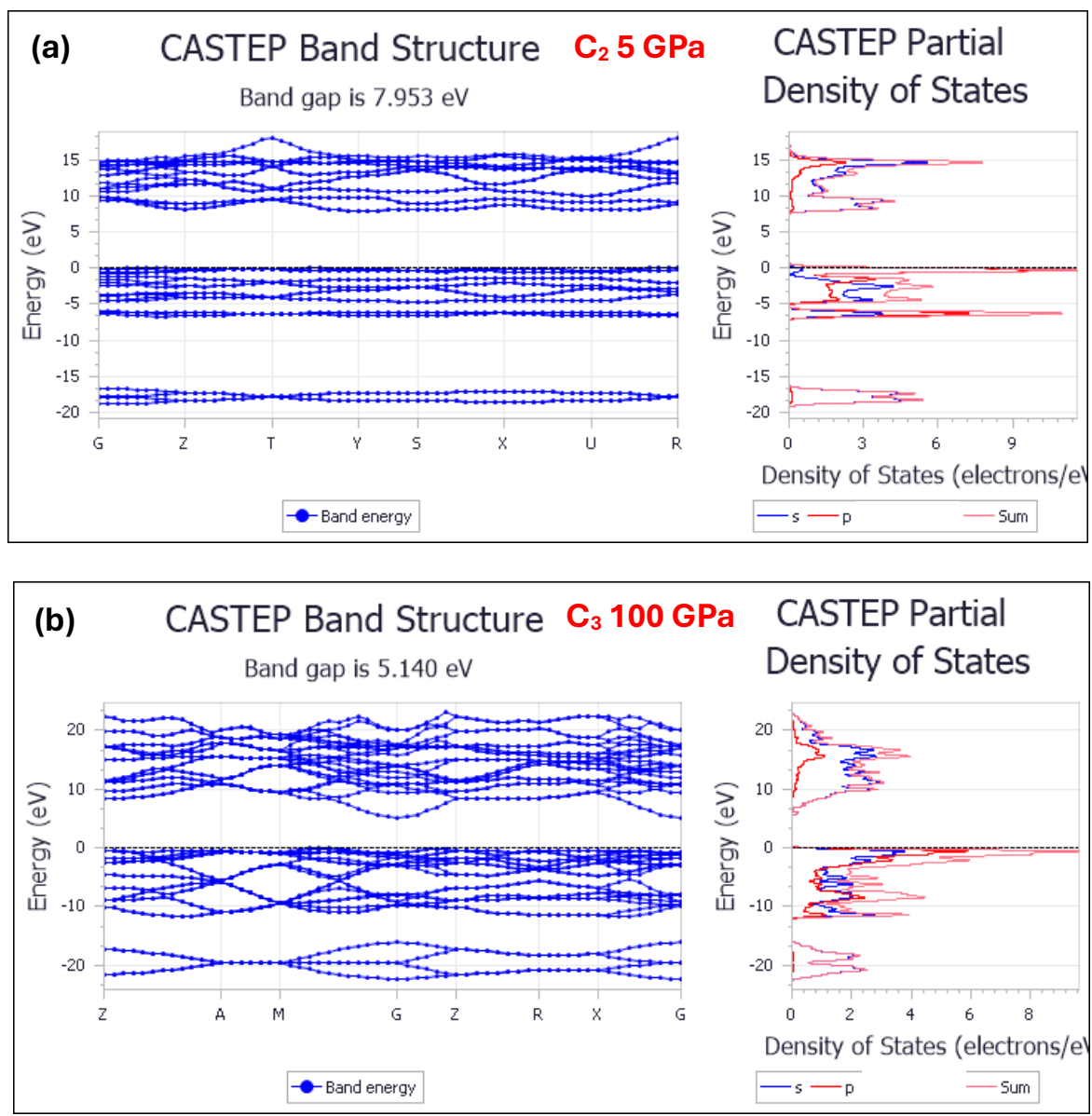

**Figure S9.** The calculated electronic band structure and electronic density of states of the $C_2$ phase at 5 GPa and the $C_3$ phase at 100 GPa.

**(a) B₁: 492 cm⁻¹**     **(b) A₁: 663 cm⁻¹**

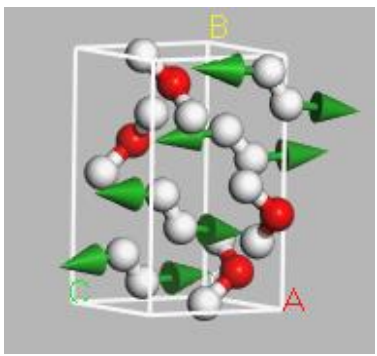 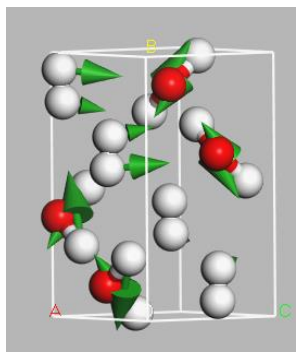

**Figure S10.** Selected normal vibrations calculated for the *Pna*2₁ C₂ phase at 40 GPa. (a) Rotational mode of $H_2$ molecules; (b) O-H stretching mode of the $H_2O$ molecules coupled with translational vibrations of the $H_2$ molecules.

**Table S1. Experimental runs.**

| Facility/date/# | Pressure (GPa) | Temperature treatments | Results |
|---|---|---|---|
| Petra III, P02.2/2016/1 | 7.5-64 | 300 K | $C_2$- $C_3$ mixed |
| Petra III, P02.2/2016/2 | 22.5-93 | 300 K | $C_2$-$C_3$ mixed |
| APS, GSECARS/2021-2023 | 31 | Near IR laser heated at 31 GPa via Au coupler | $C_2$ |
|  | 31-104 |  | $C_2$-$C_3$ mixed |
| ESRF, ID27, February 2024 | 47-50 | Near IR laser heated via Au coupler | $C_2$-$C_3$ transition |
| ESRF, ID27, April 2024 Petra III, P02.2/April 2024 | 52-64 | Near IR laser heated via Au coupler | $C_2$-$C_3$ transition |
| ESRF, ID11, July 2024 | 69 | $CO_2$ laser heated | $C_2$-$C_3$ transition |
| ESRF, ID11, July 2024 | 82-86 | $CO_2$ laser heated | $C_2$-$C_3$ transition |

**Table S2. Vibrational modes of the C$_2$ and C$_3$ phases approximated by *Pna*2$_1$ and *P*4$_1$ structures**

| Space & point group | *Pna*2$_1$ #33 (C$_{2v}$) | | *P*4$_1$ #76 (C$_4$) | |
|---|---|---|---|---|
| Site symmetry | 4a (C$_1$) | | 4a (C$_1$) | |
| Acoustic modes | A$_1$+B$_1$+B$_2$ | | A+E | |
| Optical modes | Modes | Activity | Modes | Activity |
| | 15A$_1$ | Raman & IR | 21A | Raman & IR |
| | 15A$_2$ | Raman | 21B | Raman |
| | 15B$_1$ | Raman & IR | 21E | Raman & IR |
| | 15B$_2$ | Raman & IR | | |
| H$_2$ molecular vibrons | A$_1$ | Raman & IR | 2A | Raman & IR |
| | A$_2$ | Raman | 2B | Raman |
| | B$_1$ | Raman & IR | 2E | Raman & IR |
| | B$_2$ | Raman & IR | | |